# A New Approach for Automatic Segmentation and Evaluation of Pigmentation Lesion by using Active Contour Model and Speeded Up Robust Features

Ms. Sara Mardanisamani
Department of Electrical Engineering,
Zarghan Branch, Islamic Azad University,
Zarghan, Iran

Dr. Zahra Karimi
Pharmaceutical Sciences Research Center,
Shiraz University of Medical Sciences,
Shiraz, Iran

Prof. Akram Jamshidzadeh
Department of Pharmacology and Toxicology,
Faculty of Pharmacy,
Shiraz University of Medical Sciences,
Shiraz, Iran

Prof. Mehran Yazdi
Department of Communications and Electronics,
Faculty of Electrical and Computer Engineering,
Shiraz University, Shiraz, Iran

Dr. Melika Farshad
Department of Dentistry,
Shiraz University of Medical Sciences,
Shiraz, Iran.

Mr. Amirmehdi Farshad
Department of Power and Control Engineering,
School of Electrical and Computer Engineering,
Shiraz University, Shiraz, Iran

*Abstract*— **Digital image processing techniques have wide applications in different scientific fields including the medicine. By use of image processing algorithms, physicians have been more successful in diagnosis of different diseases and have achieved much better treatment results. In this paper, we propose an automatic method for segmenting the skin lesions and extracting features that are associated to them. At this aim, a combination of Speeded-Up Robust Features (SURF) and Active Contour Model (ACM), is used. In the suggested method, at first region of skin lesion is segmented from the whole skin image, and then some features like the mean, variance, RGB and HSV parameters are extracted from the segmented region. Comparing the segmentation results, by use of Otsu thresholding, our proposed method, shows the superiority of our procedure over the Otsu theresholding method. Segmentation of the skin lesion by the proposed method and Otsu thresholding compared the results with physician's manual method. The proposed method for skin lesion segmentation, which is a combination of SURF and ACM, gives the best result. For empirical evaluation of our method, we have applied it on twenty different skin lesion images. Obtained results confirm the high performance, speed and accuracy of our method.**

I. INTRODUCTION

Quantification of skin color or determining the extent of pigmentation is necessary in dermatology and cosmetic science. In early 1920s, non-invasive techniques began to be used for skin pigmentation measurement, using different devices. Assuming that light intensity attenuates when remitted by the skin, these instruments were designed to enhance the performance of skin pigmentation by use of skin images [1].

Spectrophotometers and colorimeters, known as Skin reflectance measurement devices, supply the techniques for objective measurement and high inter/intra-rater reliability [2,3]. Not only the expense, but also needs for user training for these devices are the main limitations in their use. Therefore, developing more user-friendly and less costly methods would be advantageous for physicians and researchers.

In the last decades a number of computer vision based methods have been proposed to enhance skin images. These techniques are mostly used for melanoma diagnosis. Using Digital image process helps physicians in better diagnosis and treatments. In general physician could locate the lesions and describe their dermatological features by extracting the image parameters. Studies have shown that practically, using these automated systems is sufficient for diagnosis of skin color disorders or melanoma [4,5].

Numerous methods have been reported in computer vision and image processing, for Image segmentation [6-8]. Gabriella et al. proposed a method to detect some specific dermoscopic criteria for Melanocytic Skin Lesion which is based on blue whitish veil, the regression, and the irregular streaks techniques [9]. El-Zaart modeled a new threshold estimation method by an unsupervised learning technique with beta distribution [10].

Abbas et al. used a different approach. He rescaled all images, then reduced their artifacts by applying some filters, and finally segmented the lesion borders. This unsupervised approach uses Region-based Active Contours (RACs) for lesion border segmentation [11].

Chan kim et al. proposed a method based on L*a*b color coordinates [12] and developed a Visual Basic program to compare the pre- and post- treatment skin color. Xu et al. used another method for determination of lesion boundaries. In this method, threshold value determined for the gray-scaled images and initial lesion segmentation were achieved [8].

Using digital filters can provide efficient techniques for determining skin disorder. These methods are studied by Güçin, Patias and Altan [13].

Studying skin cancer images could be more complicated. For this matter, not only the skin color but also its texture should be used. Padmapriya Nammalwar et al. combine these two factors to present an efficient procedure for skin cancer image segmentation [14]. Fassihi et al. suggested another method for skin cancer images. In this method the segmentation is performed by morphological operators and features are extract using wavelet transforms [15].

In this paper an automatic segmentation algorithm based on SURF algorithm and Active contour model thresholding is proposed. For automatic border detection, our approach exploits two main steps: (1) finding interest point by use of the surf algorithm, (2) segmenting the skin lesions by use of the ACM. Our method is then compared with the manual segmentation results performed by an expert physician and the automatic Otsu thresholding algorithm.

## II. MATERIAL AND METHODS

### A. Image acquisition

The process of manual segmentation of the skin lesion images is very time consuming. It also requires significant medical expertise, and can be prone to error. In this work, 20 color images with 256×256 pixels, from the image database Dermatlas [16] are used for the experiments. We conducted the experiments on these skin images and compared the segmentation results obtained by our approach with those subjectively chosen by our physician and other methods. The examination is performed on 20 images, all in JPEG format. To conduct the experiments we used Matlab 7.1 on a dual core Pentium IV computer with 4GB Ram and 2.53 GHZ processor speed.

### B. Segmentation method

Segmentation is the first step of computer-based skin lesion diagnosis. The lesion boundary provides important information for accurate diagnosis. Furthermore, the extraction of other clinical features critically depends on the accuracy of the boundary [17]. Our goal in this article is to find a contour that best approximates the perimeter of a skin lesion. In this paper, we present an active contour segmentation approach and apply it to skin lesion images. For active contour segmentation, an initial contour is needed as a first step of segmentation. At this aim we use an automatic strategy, in which the initial contour is obtained roughly by Surf algorithm.

### C. Speeded Up Robust Features

For real-time visual navigation, interest point detection and description are required. SURF is one of the methods which could detect the interest point [18]. Bay et al. proposed this scale-invariant feature detector method based on Hessian-matrix [19].

Three steps involved in this method: 1- describing the basic idea of integral images, 2- obtaining approximation of the Hessian matrix and 3- computing the determinant of the Hessian matrix.

### 2.4. Interest Point Detection

An integral image I(x) at location x represents the sum of all pixels in the input image I and is defined as (1)

$$I(x) = \sum_{i=0}^{i \leq x} \sum_{j=0}^{j \leq y} I(i,j) \qquad (1)$$

By utilizing the integral image, the area within a bounded region (A, B, C, D) of the original image can be computed using four places in memory and three summations. Fig. 1 shows calculating the sum of intensities for obtaining integral images.

Hessian matrix at scale σ is defined as follows:

$$H(x, \sigma) = \begin{bmatrix} L_{xx}(x, \sigma) & L_{xy}(x, \sigma) \\ L_{xy}(x, \sigma) & L_{yy}(x, \sigma) \end{bmatrix} \qquad (2)$$

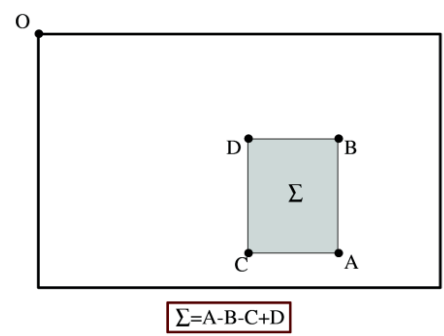

Figure 1: Calculating the sum of intensities inside a rectangular region of any size for obtaining integral images [19].

Interest points are obtained after calculating the Hessian matrix [18, 20]. These data then used to detect the keypoints. SURF first allocates an orientation to each keypoint: A circular region around the keypoint is convolved with two Haar wavelets. Scale σ, at which the keypoint is detected, determines the size of region, wavelets and sampling steps. A circular window centered on each keypoint is used for computing the SURF descriptor [21]. These keypoints are depicted in Fig. 2. After keypoints are detected with SURF algorithm, several keypoints are randomly connected to each other so that the initial contour are created.

### 2.5. Active Contour Model

To segment the skin lesion by use of ACM, interest points were extracted and used to form the initial contours automatically. Active contour is defined parametrically [22, 23, 24]. We use a type of active contour that minimizes the energy defined on contours or curves living in the domain of the image. This minimization detects specified features within an image. It is a flexible curve which can be dynamically derived to the boundary of the object of interest in the image. Fitting active contours to objects in images is an interactive process. To obtain active contour we need to have initial contour first; this could be obtained by SURF algorithm.

$$\vec{v}(s) = (\vec{x}(s), \vec{y}(s)) \quad (3)$$

Where $x(s)$ and $y(s)$ are $x$, $y$ coordinates of pixels that pass through the contour and s is the normalized index of the control points. There are two components that describe the energy function of active contour: internal and external energy. To calculate internal energy the following formula could be used:

$$E_{int} = \alpha(s)\left|\frac{dv}{ds}\right|^2 + \beta(s)\left|\frac{d^2v}{ds^2}\right|^2 \quad (4)$$

Where α is an adjustable constant that specifies continuity and β is adjustable constant that specifies contour curving. Sum of elastic and bending energies are considered as the internal energy. To calculate these two energies the following formulas are used:

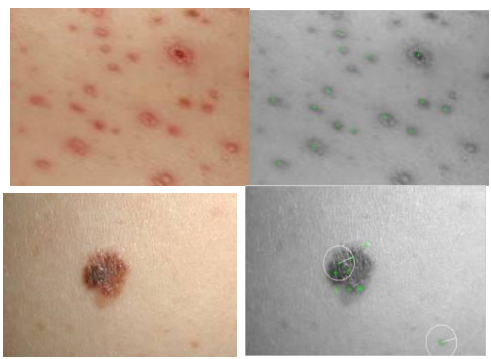

Figure 2: extracting interest point by using the SURF algorithm. The original images are colorful. Interest points are shown on gray scale images.

$$E_{elastic} = \int \alpha(\vec{v}(s) - \vec{v}(s-1))^2 ds \quad (5)$$

$$E_{bend} = \int \beta(\vec{v}(s-1) - \vec{v}(s) + \vec{v}(s+1))^2 ds \quad (6)$$

Minimized energy function is defined as follow:

$$E^*_{snake} = \int_0^1 \{E_{int}(v(s)) + E_{image}(v(s)) + E_{con}(v(s))\} ds \quad (7)$$

Where $E_{int}$, $E_{image}$ and $E_{con}$ are curve's internal energy, picture's energy and external limitations, respectively [25]. Therefore, we can successfully segment the lesions by using the SURF algorithm for interest point extraction, to form the initial contours automatically. The segmentation is then completed by using the ACM procedure. The combination of these methods gives high performance. The results of using this procedure for 20 real images are provided in the experimental section.

### 2.6. Evaluation of the Segmented Skin Lesion

After segmenting the lesion area, some features can be extracted from that area. These features can be useful for the physicians in their diagnosis and determining the degree of an especial disorder. Physicians usually use Photoshop software to obtain these features. So the automatic access to these features can be very helpful. The most useful features which could be extracted from the segmented regions are; mean and variance of the lesion area and the healthy skin part, the pixel value of images in RGB and HSV color spaces and the histogram of image in each of these two spaces. These features are depicted in Fig. 3.

### III. EXPERIMENTAL RESULTS

The images are segmented by using the active contour model and SURF method. In fact, in the images, the initial contours are obtained by the SURF algorithm. Then active contour model is used to obtain the final contours for lesions. 20 color images, with 256×256 pixels, from "Dermatlas" database are used in the experiments [16]. The obtained segmentation results were validated by an expert on this field. In Fig. 4, ten segmented images are presented: The original images (1 to 10) were processed by the proposed method and the final contours presented in the images (a to l, respectively) were obtained. All final contours were visually evaluated by an expert. He confirmed that all ill regions were successfully detected. After segmenting the skin lesion areas from the images, in our work, we extracted some features like the mean, variance, RGB and HSV parameters from the segmented regions [26] features was shown in Fig. 3.

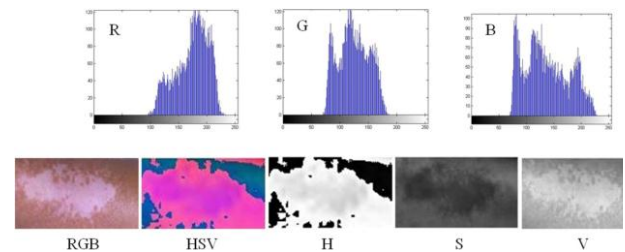

Figure 3. Top pictures: Histogram of RGB space components. Bottom pictures: Image in RGB and HSV color spaces and HSV components of the image.

We apply the Otsu's method on each box containing a skin lesion. To do this, a four class thresholding is used. Some examples of the results obtained by applying this method are shown in Fig. 5.

The results of segmenting the skin lesion, based on our method in comparison with manual physician segmentation and Otsu thresholding method are given in Table 1 and Table 2 for 20 images. The manual segmentations provided by our physician are used for performance evaluation. Table 3, shows the average recall and precision values in segmenting les for all the images. The first row in this table shows the percentage average recall and precision of segmentation, achieved in our method with respect to manual segmentation by physicians. The second row shows those results for the Otsu thresholding method with respect to the manual segmentation. Results indicate that our method have good performance in comparison with Otsu procedure. Recall and precision rates are computed here by use of relations (8) and (9), [27].

TPs indicates the skin lesion pixels correctly classified. FPs represents the healthy skin pixels classified as skin lesion pixels. TNs indicates the healthy pixels correctly classified. And finally FNs shows the skin lesion pixels classified as healthy skin pixels. The following measures have been applied to evaluate the algorithms [28].

Recall: The ratio between the number of skin lesion pixels correctly classified and the total number of actual skin lesion pixels (TPr)

$$\text{Recall} = \frac{TP}{TP+FN} \qquad (8)$$

Precision: The ratio between the number of skin lesion pixels correctly classified and the total number of pixels labeled as skin lesion pixels, by the applied skin lesion segmentation method.

$$\text{Precision} = \frac{TP}{TP+FP} \qquad (9)$$

## IV. CONCLUSION

The proposed image segmentation method uses the active contour model and SURF algorithm. As active contours always provide continuous boundaries of sub-regions, they can produce more reasonable segmentation results than traditional segmentation methods, and consequently improve the final results of image analysis. The mathematical implementation of the proposed active contour models is accomplished using level set method. By presenting contours as a level of a topological function, we can merge multiple contours into one contour, or can split a contour into multiple contours, providing a good flexibility in the use of active contours. The proposed image segmentation method in this work is successfully used for detection of lesions in real skin images.

The proposed algorithm demonstrate good performance for segmenting skin lesion and extracting features such as variance and mean. These feature can help physician to diagnosis and treat skin diseases. We do this using a fully automatic and accurate approach. Many physician obtain these features with non-automatic methods such as using Photoshop software that takes a much time.

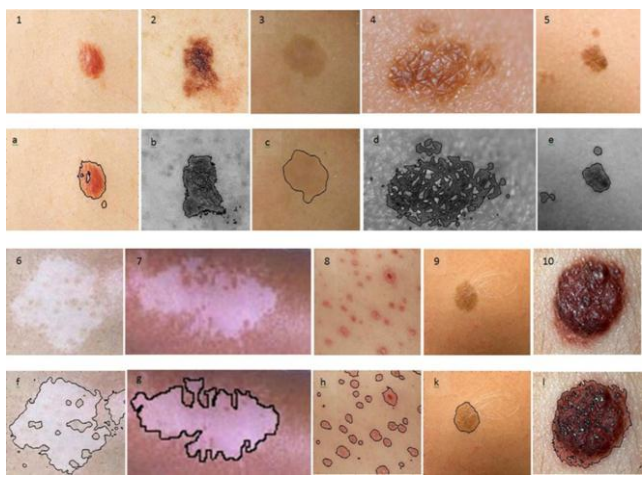

Figure 4. The results of segmenting the skin lesions in 10 images, by drawing the lesions' contours based on the proposed method.

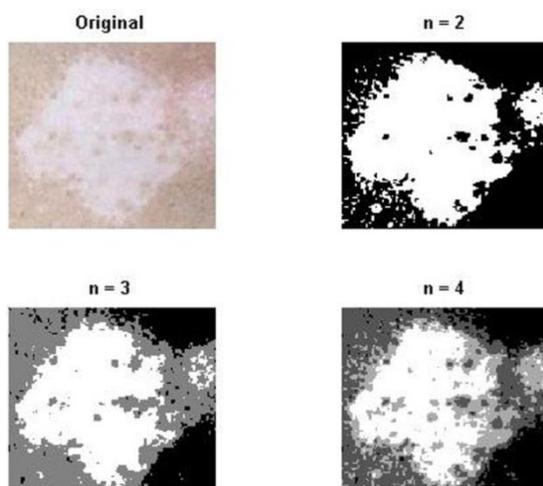

Figure 5. Results of skin lesion segmentation by using Otsu's thresholding method in four classes.

TABLE 1 . SEGMENTATION OF THE SKIN LESION BY THE PROPOSED METHOD COMPARED THE RESULTS WITH PHYSICIAN'S MANUAL METHOD BY USE OF RECALL AND PRECISION FACTORS

| Skin image number | 1 | 2 | 3 | 4 | 5 | 6 | 7 | 8 | 9 | 10 |
|---|---|---|---|---|---|---|---|---|---|---|
| Recall | 0.96 | 0.91 | 0.87 | 0.95 | 0.93 | 0.90 | 0.92 | 0.92 | 0.88 | 0.93 |
| Precision | 0.82 | 0.94 | 0.92 | 0.95 | 0.98 | 0.93 | 0.94 | 0.82 | 0.97 | 0.95 |
| Skin image number | 11 | 12 | 13 | 14 | 15 | 16 | 17 | 18 | 19 | 20 |
| Recall | 0.96 | 0.88 | 0.80 | 0.93 | 0.82 | 0.91 | 0.91 | 0.88 | 0.94 | 0.98 |
| Precision | 0.92 | 0.99 | 1 | 0.98 | 0.77 | 0.76 | 0.90 | 0.94 | 0.96 | 0.93 |

TABLE 2 . SEGMENTATION OF THE SKIN LESION BY OTSU THRESHOLDING METHOD COMPARED THE RESULTS WITH RESULTS WITH PHYSICIAN'S MANUAL METHOD BY USE OF RECALL AND PRECISION FACTORS

| Skin image number | 1 | 2 | 3 | 4 | 5 | 6 | 7 | 8 | 9 | 10 |
|---|---|---|---|---|---|---|---|---|---|---|
| Recall | 0.98 | 0.74 | 0.97 | 0.97 | 0.75 | 0.14 | 0.14 | 0.75 | 0.67 | 0.96 |
| Precision | 0.06 | 0.98 | 0.82 | 0.27 | 0.8 | 0.1 | 0.14 | 0.91 | 0.99 | 0.26 |
| Skin image number | 11 | 12 | 13 | 14 | 15 | 16 | 17 | 18 | 19 | 20 |
| Recall | 0.97 | 0.79 | 0.73 | 0.68 | 0.63 | 0.98 | 0.75 | 0.75 | 0.88 | 0.78 |
| Precision | 0.14 | 0.99 | 0.89 | 0.51 | 0.35 | 0.015 | 0.92 | 0.99 | 0.43 | 0.18 |

TABLE 1 . COMPARING RESULT BETWEEN OUR METHOD AND OTSU THRESHOLDING ALGORITHM BY USE OF RECALL AND PRECISION FACTORS

|  | Average Recall Percentage | Average Precision Percentage |
|---|---|---|
| Our method | 91 | 92 |
| Otsu thresholding method | 75 | 54 |

ACKNOWLEGMENT

The result of this investigation was originated from the research project funded by research affairs office of Islamic Azad University, Zarghan Branch. The authors express gratitude to Islamic Azad University Zarghan Branch because of financial support.